\documentclass[]{ceurart}

\usepackage{listings}
\usepackage{graphics}
\usepackage{adjustbox}
\usepackage{wrapfig}
\usepackage{url}
\usepackage{graphicx}
\usepackage{float,wrapfig}
\usepackage{xcolor}

\begin{document}

\copyrightyear{2024}
\copyrightclause{Copyright for this paper by its authors.
  Use permitted under Creative Commons License Attribution 4.0
  International (CC BY 4.0).}

\conference{ISWC'24: The 23rd International Semantic Web Conference,
  November 11--15, 2024, Baltimore, USA}

\title{Towards a Knowledge Graph for Teaching Knowledge Graphs}

% \tnotetext[1]{You can use this document as the template for preparing your
%   publication. We recommend using the latest version of the ceurart style.}

\author[1]{Eleni Ilkou}[%
orcid=0000-0002-4847-6177,
email=ilkou.el@gmail.com,
]
\cormark[1]
\fnmark[1]
% \address[1]{Peoples' Friendship University of Russia (RUDN University),
%   6 Miklukho-Maklaya St, Moscow, 117198, Russian Federation}
% \address[2]{Joint Institute for Nuclear Research,
%   6 Joliot-Curie, Dubna, Moscow region, 141980, Russian Federation}

% Ernesto please add your data :)
\author[2,3]{Ernesto Jim\'enez-Ruiz}[
 orcid=0000-0002-9083-4599,
 email=ernesto.jimenez-ruiz@city.ac.uk,
 %url=https://www.city.ac.uk/about/people/academics/ernesto-jimenez-ruiz,
]
\address[1]{L3S Research Center, Leibniz University Hannover, Germany}
 \address[2]{City St George's, University of London, UK}
 \address[3]{University of Oslo, Norway}

% \author[4]{Manfred Jeusfeld}[%
% orcid=0000-0002-9421-8566,
% email=Manfred.Jeusfeld@acm.org,
% url=http://conceptbase.sourceforge.net/mjf/,
% \address[4]{University of Skövde, Högskolevägen 1, 541 28 Skövde, Sweden}

\cortext[1]{Corresponding author.}
\fntext[1]{This work was performed while a research visitor at City, University of London}

\begin{abstract}
This poster paper describes the ongoing research project for the creation of a use-case-driven Knowledge Graph resource tailored to the needs of teaching education in Knowledge Graphs (KGs).
We gather resources related to KG courses from lectures offered by the Semantic Web community, with the help of the COST Action Distributed Knowledge Graphs and the interest group on KGs at The Alan Turing Institute.
Our goal is to create a resource-focused KG with multiple interconnected semantic layers that interlink topics, courses, and materials with each lecturer. Our approach formulates a domain KG in teaching and relates it with multiple Personal KGs created for the lecturers. \href{https://naiayti.github.io/TeachingKnowledgeGraphs.io/}{Poster Page (GitHub): https://naiayti.github.io/TeachingKnowledgeGraphs.io/}

\end{abstract}

\begin{keywords}
  Educational Knowledge Graph \sep
  Domain Model \sep
  Teacher Model \sep
  Personal Knowledge Graph
\end{keywords}

\maketitle

\vspace{-6mm}
\section{You ask, we answer: an unorthodox Introduction}

\paragraph{Is there really a problem?}
Teaching courses at the university level can be a demanding task; the field is radically evolving with topic trends changing each year, the material needs to be up-to-date, and finding teaching resources for a topic outside of one's main expertise can be challenging. The challenge arises both from not having a central hub, which can assist in accessing educational resources, topics, and courses, and also from the pluralistic naming and nonuniform format of similar resources across different institutions. Moreover, in the courses offered by the semantic web community, we observe that equivalant courses cover diverse topics, and these courses often highly differ in naming, such as ``Web AI'' and ``Knowledge Engineering''. Also, the educational material for lectures and labs is found in a large variety of modalities. 
From the COST DKG Workshop ``Teaching Knowledge Graphs", Malaga, September 2023, it was identified the necessity for providing a resource that could highlight the topics and material taught in KG courses and a platform in which lectures could connect and exchange materials. Therefore, the proposal for a teaching Knowledge Graph (KG) resource was highlighted.

\vspace{-3mm}
\paragraph{Who cares?}
The primary beneficiaries are the lecturers who teach Knowledge Graph (KG) courses, and students who follow these courses. Those can benefit from a teaching KG in multiple use cases, as indicated in Table~\ref{tab: the CQs}, as
the resource is designed with the intention of supporting teaching and learning in the discovery of related topics, courses, datasets and educational materials related to KGs.
By enhancing the teaching process, the lecturers can leverage the resource to better explain sub-concepts, demonstrate practical applications, and provide richer educational experiences. 
Additionally, the students' learning is enhanced as they can have access to a central hub with well-structured information about KG courses, and leverage similar content to their course to gain a broader perspective about their study. 
Subsequently, the teaching KG resource can function as a digital library with content quality assurance, where educational materials are organized by level and topic, and interconnected with the lecturers.

\begin{wraptable}{}{7.33cm}
\vspace{-6.5mm}

\caption{The use cases presented as competency questions }

\begin{tabular}{c}
\toprule
\textbf{Competency Questions }             \\
\midrule
For (sub)topic X                 \\ 
\midrule
Who is teaching X?                \\
Which are the materials for X?    \\
Which are the prerequisites of X? \\
Which are the labs for X?   \\ 
\midrule
For course Y                      \\ 
\midrule
Who is the target audience for Y? \\
Which educational level does Y target?                    \\
Who is teaching Y?                \\
Which slides are linked to Y?     \\
Which labs are part of Y?         \\
Which courses are similar to Y?   \\
Which are suggested resources for Y?                            \\ \hline
For dataset D                     \\ \hline
Which exercises exist for D?      \\
Which courses use D?              \\ \hline
For material M                    \\ \hline
Which courses use a material similar to M?                     \\
How much similar is M to another material?              \\
Which topics does M cover?        \\
Is M open access?                 \\ \hline
\end{tabular}

\label{tab: the CQs}
\vspace{-3mm}
\end{wraptable}

Moreover, the Semantic Web (SW) community needs a methodology for creating an educational teaching KG. This need is two folded; at first, a methodology generated from the community  will benefit educators and practitioners. A SW framework for creating educational teaching KGs will grant educators access to the latest advancements that utilize the state-of-the-art standards, and fully explore the capabilities offered by the SW.
On the other hand, as education is evolving, the demand for semantic solutions is prominent highlighting the need for the involvement of the SW into the theoretical foundation of learning and teaching applications.

\vspace{-3mm}
\paragraph{Ok, but isn't it already done?}

All we know, we are the first working towards a teaching KG for KG courses. However, our contribution does not rely only on this. 
As Semantic Web technologies have been widely used in the education domain, ontologies are commonly used for semantically enhancing e-learning applications~\cite{DBLP:conf/semweb/IlkouATHKAN21, DBLP:journals/caeai/RahayuFK22}.
Additionally, educational ontologies have been deployed for KG construction~\cite{hubertISWC2022} and recently, a few educational-purpose KGs have been created for a plethora of applications~\cite{abu2024systematic,ilkou2020technology, DBLP:journals/access/ChenLZCY18,dang2019mooc,shen2021ckgg, DBLP:conf/ksem/ChenYFQRX21, DBLP:conf/icalt/AbuRasheedDWKBF23, DBLP:conf/ectel/AbuRasheedWDF23, DBLP:journals/jodl/KabongoDA24}. 
Moreover, Personal Knowledge Graphs (PKGs) have recently been developed for educational applications~\cite{DBLP:conf/lak/AinCKAJS24, DBLP:conf/www/Ilkou22, DBLP:journals/debu/ChakrabortyDSMD23}, where they utilise semantic resources, such as linking entities to encyclopedic KGs. Nonetheless, they are not used in connection to a bigger domain or encyclopedic KG as they do not exchange information, such as updating information about the triples in the encyclopedic or domain KG.

However, none of these approaches goes beyond the factual knowledge representation and limits each resource description to high-level connections, such as which resource is taught first and who is the resource's author.
To the best of our knowledge, we are the first to aim to extract knowledge from each resource, such as topics from educational material, and represent them as new knowledge in the KG.
This is significantly important, as one of our main contributions lies in defining a new way of creating educational KGs. We achieve this by building a use-case-driven and resource-focused KG that offers a novel representation of the data it contains. The new representation goes beyond the factual knowledge and extraction of metadata, and further includes the statistical analysis and document information retrieval outcomes as part of the KG entities and properties. An example of this case is applying topic modelling to lecture notes~\cite{vayansky2020review}, and enhancing the teaching KG with new entities as subtopics extracted from the lecture notes.

\vspace{-3mm}
\paragraph{Is there a big picture or you just want a paper published?}

Our effort is a stepping stone towards a big, multi-lingual, multi-modal, and high quality content educational teaching KG that can serve as the basis of future educational applications in AI and beyond~\cite{ilkou2023edumultikg}. As KGs can assist AI applications in education to offer more personalised and better learning experience to the learners~\cite{abu2024systematic}, our goal is to enhance teachers experience by creating the basis for educational teaching KGs focused on quality content.

Also, as quality education is underlined as a Sustainability Goal~\cite{SDG4}, raises the duty of setting standards for educational Semantic Web applications and KG resources in education. Consequently, we guarantee the quality of our resource by gathering topic descriptions and courses from the experts who teach them.

\vspace{-3mm}
\paragraph{Fine, but how are you going to make it?}

We reuse Semantic Web vocabularies, and standards to develop the teaching KG~\cite{DBLP:journals/eaai/Poveda-Villalon22}.
Our resource aims to interlink topics, courses, and educational material in order to enable teachers to enhance their courses and help students access multiple resources relevant to their learning goals. Driven by the needs of the users and adding granularity to the representation of material in our teaching KG resource, we aim to contribute both to the Semantic Web advancements as well as set the basis for advanced learning analytics applications.

The teaching KG consists of two parts, the domain model and the user model, which are developed as extensions of the EduCOR ontology~\cite{DBLP:conf/semweb/IlkouATHKAN21} and PKGs.
At first, we create a domain model, the teaching KG resource, which extends the EduCOR ontology. The domain model contains the topics taught, the educational material, such as the lecture notes and labs, and the courses as sequences of the material. Moreover, we model each lecturer as a user model via PKGs~\cite{DBLP:conf/www/Ilkou22} which have been shown capable of enabling smart learning environment analytics~\cite{DBLP:journals/tlt/IlkouTFT23}. Each user model contains the lecturer identification information and is connected with the lecturer course in the domain model.

\paragraph{Aren't there any limitations?}

There are plenty of limitations to consider when building teaching KGs.
Firstly, not all materials are open source and accessible, as frequently educational content stays hidden behind paywalls or requires institutional access. To address this, we develop the PKGs so the users can access the specific materials from each instructor and gain access to that knowledge. Moreover, the diversity in educational material and non-standardized course descriptions make automation the biggest challenge. The material and courses are present in various formats, which make their integration, interlinking, and maintenance a non-trivial task that we aim to tackle by parsing the text that can be extracted from each resource using symbolic and sub-symbolic techniques~\cite{ilkou2020symbolic}.

\vspace{-4mm}
\section{Can you give me an example?}
\vspace{-2mm}

\begin{wrapfigure}{r}{9cm}
\vspace{-5mm}
    \centering
    % Answer: [trim={left bottom right top},clip]
    \includegraphics[trim={0.76cm 0.8cm 0.76cm 1cm},  clip, width=8.5cm]{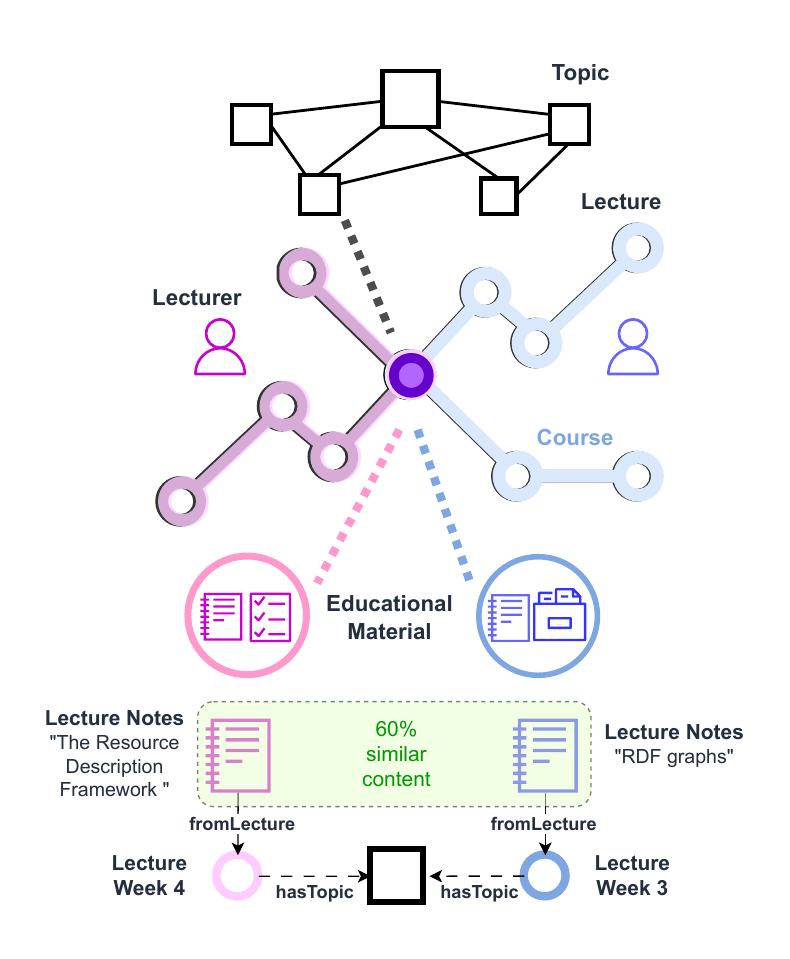}
    \vspace{-2mm}    
    \caption{A visualisation of the teaching KG components (top) and application (bottom). On the top is the topic ontology, which contains topics and sub-topics (squares). The topic ontology is interconnected with week lectures (circles), which are part of a course (circles and edges in a single color). Each course has a lecturer (human icon). Each lecture has one or many educational material (the content in the circles).}
    \label{fig:enter-label}
\end{wrapfigure}

In Figure~\ref{fig:enter-label}, we provide a visualisation of the multiple layers (topic, lecture, lecturer, course, educational material) present in the teaching KG, in the top part of the Figure, and a paradigm for assisting content retrieval based on semantic similarity, in the bottom part. 
For the latter, we follow techniques similar to literature~\cite{DBLP:conf/icalt/AbuRasheedDWKBF23, DBLP:conf/icalt/LiCZZZ23} to compute the semantic similarity between resources 
which enables to classify and access similar resources via the connections to the topics ontology. Therefore, by interconnecting the lectures with topics we can compare the educational material they contain, and determine which lectures are similar under the same topic.

Moreover, we foresee a hierarchical structure in the topics level. As topics can be classified to general as hypernyms, such as ``Data Representation'', and more specified as subtopics, such as ``RDFS representation''. We can introduce the relation of prerequisite among topics via statistical analysis of the available resources. The prerequisites are extracted by the course order of lectures and the topics those materials cover. For example, the topic of RDF based on statistical analysis of the gathered material it would be a prerequisite of the SPARQL topic.


% % Add in camera ready
\begin{acknowledgments}
\vspace{-2mm}
We thank Axel Polleres for the name inspiration. This work was supported by COST Action Distributed Knowledge Graphs and the interest group on Knowledge Graphs at The Alan Turing Institute. E.I. would like to thank her grandma for the eternal love.
\end{acknowledgments}
% \newpage

%%
%% Define the bibliography file to be used
\bibliography{sample-ceur}

\begin{thebibliography}{22}
\expandafter\ifx\csname natexlab\endcsname\relax\def\natexlab#1{#1}\fi
\providecommand{\url}[1]{\texttt{#1}}
\providecommand{\href}[2]{#2}
\providecommand{\path}[1]{#1}
\providecommand{\DOIprefix}{doi:}
\providecommand{\ArXivprefix}{arXiv:}
\providecommand{\URLprefix}{URL: }
\providecommand{\Pubmedprefix}{pmid:}
\providecommand{\doi}[1]{\href{http://dx.doi.org/#1}{\path{#1}}}
\providecommand{\Pubmed}[1]{\href{pmid:#1}{\path{#1}}}
\providecommand{\bibinfo}[2]{#2}
\ifx\xfnm\relax \def\xfnm[#1]{\unskip,\space#1}\fi
%Type = Inproceedings
\bibitem[{Ilkou et~al.(2021)Ilkou, Abu{-}Rasheed, Tavakoli, Hakimov, Kismih{\'{o}}k, Auer, and Nejdl}]{DBLP:conf/semweb/IlkouATHKAN21}
\bibinfo{author}{E.~Ilkou}, \bibinfo{author}{H.~Abu{-}Rasheed}, \bibinfo{author}{M.~Tavakoli}, \bibinfo{author}{S.~Hakimov}, \bibinfo{author}{G.~Kismih{\'{o}}k}, \bibinfo{author}{S.~Auer}, \bibinfo{author}{W.~Nejdl},
\newblock \bibinfo{title}{Educor: An educational and career-oriented recommendation ontology},
\newblock in: \bibinfo{booktitle}{The Semantic Web - {ISWC} 2021 - 20th International Semantic Web Conference, {ISWC} 2021, Virtual Event, October 24-28, 2021, Proceedings}, volume \bibinfo{volume}{12922} of \textit{\bibinfo{series}{Lecture Notes in Computer Science}}, \bibinfo{publisher}{Springer}, \bibinfo{year}{2021}, pp. \bibinfo{pages}{546--562}. \URLprefix \url{https://doi.org/10.1007/978-3-030-88361-4\_32}. \DOIprefix\doi{10.1007/978-3-030-88361-4\_32}.
%Type = Article
\bibitem[{Rahayu et~al.(2022)Rahayu, Ferdiana, and Kusumawardani}]{DBLP:journals/caeai/RahayuFK22}
\bibinfo{author}{N.~W. Rahayu}, \bibinfo{author}{R.~Ferdiana}, \bibinfo{author}{S.~S. Kusumawardani},
\newblock \bibinfo{title}{A systematic review of ontology use in e-learning recommender system},
\newblock \bibinfo{journal}{Comput. Educ. Artif. Intell.} \bibinfo{volume}{3} (\bibinfo{year}{2022}) \bibinfo{pages}{100047}. \URLprefix \url{https://doi.org/10.1016/j.caeai.2022.100047}. \DOIprefix\doi{10.1016/J.CAEAI.2022.100047}.
%Type = Inproceedings
\bibitem[{Hubert et~al.(2022)Hubert, Brun, and Monticolo}]{hubertISWC2022}
\bibinfo{author}{N.~Hubert}, \bibinfo{author}{A.~Brun}, \bibinfo{author}{D.~Monticolo},
\newblock \bibinfo{title}{New ontology and knowledge graph for university curriculum recommendation},
\newblock in: \bibinfo{booktitle}{Proceedings of the {ISWC} 2022 Posters, Demos and Industry Tracks: From Novel Ideas to Industrial Practice co-located with 21st International Semantic Web Conference {(ISWC} 2022), Virtual Conference, Hangzhou, China, October 23-27, 2022}, volume \bibinfo{volume}{3254} of \textit{\bibinfo{series}{{CEUR} Workshop Proceedings}}, \bibinfo{publisher}{CEUR-WS.org}, \bibinfo{address}{Online}, \bibinfo{year}{2022}. \URLprefix \url{http://ceur-ws.org/Vol-3254/paper349.pdf}.
%Type = Article
\bibitem[{Abu-Salih and Alotaibi(2024)}]{abu2024systematic}
\bibinfo{author}{B.~Abu-Salih}, \bibinfo{author}{S.~Alotaibi},
\newblock \bibinfo{title}{A systematic literature review of knowledge graph construction and application in education},
\newblock \bibinfo{journal}{Heliyon}  (\bibinfo{year}{2024}).
%Type = Inproceedings
\bibitem[{Ilkou and Signer(2020)}]{ilkou2020technology}
\bibinfo{author}{E.~Ilkou}, \bibinfo{author}{B.~Signer},
\newblock \bibinfo{title}{A technology-enhanced smart learning environment based on the combination of knowledge graphs and learning paths.},
\newblock in: \bibinfo{booktitle}{CSEDU (2)}, \bibinfo{year}{2020}, pp. \bibinfo{pages}{461--468}.
%Type = Article
\bibitem[{Chen et~al.(2018)Chen, Lu, Zheng, Chen, and Yang}]{DBLP:journals/access/ChenLZCY18}
\bibinfo{author}{P.~Chen}, \bibinfo{author}{Y.~Lu}, \bibinfo{author}{V.~W. Zheng}, \bibinfo{author}{X.~Chen}, \bibinfo{author}{B.~Yang},
\newblock \bibinfo{title}{Knowedu: {A} system to construct knowledge graph for education},
\newblock \bibinfo{journal}{{IEEE} Access} \bibinfo{volume}{6} (\bibinfo{year}{2018}) \bibinfo{pages}{31553--31563}. \URLprefix \url{https://doi.org/10.1109/ACCESS.2018.2839607}. \DOIprefix\doi{10.1109/ACCESS.2018.2839607}.
%Type = Inproceedings
\bibitem[{Dang et~al.(2019)Dang, Tang, and Li}]{dang2019mooc}
\bibinfo{author}{F.~Dang}, \bibinfo{author}{J.~Tang}, \bibinfo{author}{S.~Li},
\newblock \bibinfo{title}{Mooc-kg: a mooc knowledge graph for cross-platform online learning resources},
\newblock in: \bibinfo{booktitle}{2019 IEEE 9th International Conference on Electronics Information and Emergency Communication (ICEIEC)}, \bibinfo{organization}{IEEE}, \bibinfo{year}{2019}, pp. \bibinfo{pages}{1--8}. \URLprefix \url{https://ieeexplore.ieee.org/stamp/stamp.jsp?arnumber=8784572}.
%Type = Inproceedings
\bibitem[{Shen et~al.(2021)Shen, Chen, Cheng, and Qu}]{shen2021ckgg}
\bibinfo{author}{Y.~Shen}, \bibinfo{author}{Z.~Chen}, \bibinfo{author}{G.~Cheng}, \bibinfo{author}{Y.~Qu},
\newblock \bibinfo{title}{{CKGG:} {A} chinese knowledge graph for high-school geography education and beyond},
\newblock in: \bibinfo{booktitle}{The Semantic Web - {ISWC} 2021 - 20th International Semantic Web Conference, {ISWC} 2021, Virtual Event, October 24-28, 2021, Proceedings}, volume \bibinfo{volume}{12922} of \textit{\bibinfo{series}{Lecture Notes in Computer Science}}, \bibinfo{publisher}{Springer}, \bibinfo{address}{Online}, \bibinfo{year}{2021}, pp. \bibinfo{pages}{429--445}. \URLprefix \url{https://doi.org/10.1007/978-3-030-88361-4\_25}. \DOIprefix\doi{10.1007/978-3-030-88361-4\_25}.
%Type = Inproceedings
\bibitem[{Chen et~al.(2021)Chen, Yin, Fan, Qiao, Rong, and Xiong}]{DBLP:conf/ksem/ChenYFQRX21}
\bibinfo{author}{H.~Chen}, \bibinfo{author}{C.~Yin}, \bibinfo{author}{X.~Fan}, \bibinfo{author}{L.~Qiao}, \bibinfo{author}{W.~Rong}, \bibinfo{author}{Z.~Xiong},
\newblock \bibinfo{title}{Learning path recommendation for {MOOC} platforms based on a knowledge graph},
\newblock in: \bibinfo{booktitle}{Knowledge Science, Engineering and Management - 14th International Conference, {KSEM} 2021, Tokyo, Japan, August 14-16, 2021, Proceedings, Part {II}}, volume \bibinfo{volume}{12816} of \textit{\bibinfo{series}{Lecture Notes in Computer Science}}, \bibinfo{publisher}{Springer}, \bibinfo{year}{2021}, pp. \bibinfo{pages}{600--611}. \URLprefix \url{https://doi.org/10.1007/978-3-030-82147-0\_49}. \DOIprefix\doi{10.1007/978-3-030-82147-0\_49}.
%Type = Inproceedings
\bibitem[{Abu{-}Rasheed et~al.(2023{\natexlab{a}})Abu{-}Rasheed, Dornh{\"{o}}fer, Weber, Kismih{\'{o}}k, Buchmann, and Fathi}]{DBLP:conf/icalt/AbuRasheedDWKBF23}
\bibinfo{author}{H.~Abu{-}Rasheed}, \bibinfo{author}{M.~Dornh{\"{o}}fer}, \bibinfo{author}{C.~Weber}, \bibinfo{author}{G.~Kismih{\'{o}}k}, \bibinfo{author}{U.~Buchmann}, \bibinfo{author}{M.~Fathi},
\newblock \bibinfo{title}{Building contextual knowledge graphs for personalized learning recommendations using text mining and semantic graph completion},
\newblock in: \bibinfo{booktitle}{{IEEE} International Conference on Advanced Learning Technologies, {ICALT} 2023, Orem, UT, USA, July 10-13, 2023}, \bibinfo{publisher}{{IEEE}}, \bibinfo{year}{2023}{\natexlab{a}}, pp. \bibinfo{pages}{36--40}. \URLprefix \url{https://doi.org/10.1109/ICALT58122.2023.00016}. \DOIprefix\doi{10.1109/ICALT58122.2023.00016}.
%Type = Inproceedings
\bibitem[{Abu{-}Rasheed et~al.(2023{\natexlab{b}})Abu{-}Rasheed, Weber, Dornh{\"{o}}fer, and Fathi}]{DBLP:conf/ectel/AbuRasheedWDF23}
\bibinfo{author}{H.~Abu{-}Rasheed}, \bibinfo{author}{C.~Weber}, \bibinfo{author}{M.~Dornh{\"{o}}fer}, \bibinfo{author}{M.~Fathi},
\newblock \bibinfo{title}{Pedagogically-informed implementation of reinforcement learning on knowledge graphs for context-aware learning recommendations},
\newblock in: \bibinfo{booktitle}{Responsive and Sustainable Educational Futures - 18th European Conference on Technology Enhanced Learning, {EC-TEL} 2023, Aveiro, Portugal, September 4-8, 2023, Proceedings}, volume \bibinfo{volume}{14200} of \textit{\bibinfo{series}{Lecture Notes in Computer Science}}, \bibinfo{publisher}{Springer}, \bibinfo{year}{2023}{\natexlab{b}}, pp. \bibinfo{pages}{518--523}. \URLprefix \url{https://doi.org/10.1007/978-3-031-42682-7\_35}. \DOIprefix\doi{10.1007/978-3-031-42682-7\_35}.
%Type = Article
\bibitem[{Kabongo et~al.(2024)Kabongo, D'Souza, and Auer}]{DBLP:journals/jodl/KabongoDA24}
\bibinfo{author}{S.~Kabongo}, \bibinfo{author}{J.~D'Souza}, \bibinfo{author}{S.~Auer},
\newblock \bibinfo{title}{Orkg-leaderboards: a systematic workflow for mining leaderboards as a knowledge graph},
\newblock \bibinfo{journal}{Int. J. Digit. Libr.} \bibinfo{volume}{25} (\bibinfo{year}{2024}) \bibinfo{pages}{41--54}. \URLprefix \url{https://doi.org/10.1007/s00799-023-00366-1}. \DOIprefix\doi{10.1007/S00799-023-00366-1}.
%Type = Inproceedings
\bibitem[{Ain et~al.(2024)Ain, Chatti, Kamdem, Alatrash, Joarder, and Siepmann}]{DBLP:conf/lak/AinCKAJS24}
\bibinfo{author}{Q.~U. Ain}, \bibinfo{author}{M.~A. Chatti}, \bibinfo{author}{P.~A.~M. Kamdem}, \bibinfo{author}{R.~Alatrash}, \bibinfo{author}{S.~A. Joarder}, \bibinfo{author}{C.~Siepmann},
\newblock \bibinfo{title}{Learner modeling and recommendation of learning resources using personal knowledge graphs},
\newblock in: \bibinfo{booktitle}{Proceedings of the 14th Learning Analytics and Knowledge Conference, {LAK} 2024, Kyoto, Japan, March 18-22, 2024}, \bibinfo{publisher}{{ACM}}, \bibinfo{year}{2024}, pp. \bibinfo{pages}{273--283}. \URLprefix \url{https://doi.org/10.1145/3636555.3636881}. \DOIprefix\doi{10.1145/3636555.3636881}.
%Type = Inproceedings
\bibitem[{Ilkou(2022)}]{DBLP:conf/www/Ilkou22}
\bibinfo{author}{E.~Ilkou},
\newblock \bibinfo{title}{Personal knowledge graphs: Use cases in e-learning platforms},
\newblock in: \bibinfo{booktitle}{Companion of The Web Conference 2022, Virtual Event / Lyon, France, April 25 - 29, 2022}, \bibinfo{publisher}{{ACM}}, \bibinfo{year}{2022}, pp. \bibinfo{pages}{344--348}. \URLprefix \url{https://doi.org/10.1145/3487553.3524196}.
%Type = Article
\bibitem[{Chakraborty et~al.(2023)Chakraborty, Dutta, Sanyal, Majumdar, and Das}]{DBLP:journals/debu/ChakrabortyDSMD23}
\bibinfo{author}{P.~Chakraborty}, \bibinfo{author}{S.~Dutta}, \bibinfo{author}{D.~K. Sanyal}, \bibinfo{author}{S.~Majumdar}, \bibinfo{author}{P.~P. Das},
\newblock \bibinfo{title}{Bringing order to chaos: Conceptualizing a personal research knowledge graph for scientists},
\newblock \bibinfo{journal}{{IEEE} Data Eng. Bull.} \bibinfo{volume}{46} (\bibinfo{year}{2023}) \bibinfo{pages}{43--56}. \URLprefix \url{http://sites.computer.org/debull/A23dec/p43.pdf}.
%Type = Article
\bibitem[{Vayansky and Kumar(2020)}]{vayansky2020review}
\bibinfo{author}{I.~Vayansky}, \bibinfo{author}{S.~A. Kumar},
\newblock \bibinfo{title}{A review of topic modeling methods},
\newblock \bibinfo{journal}{Information Systems} \bibinfo{volume}{94} (\bibinfo{year}{2020}) \bibinfo{pages}{101582}.
%Type = Inproceedings
\bibitem[{Ilkou et~al.(2023)Ilkou, Galletti, and Dobriy}]{ilkou2023edumultikg}
\bibinfo{author}{E.~Ilkou}, \bibinfo{author}{M.~Galletti}, \bibinfo{author}{D.~Dobriy},
\newblock \bibinfo{title}{Edumultikg attains 92\% accuracy in k-12 user profiling!},
\newblock in: \bibinfo{booktitle}{Proceedings of the ESWC}, volume \bibinfo{volume}{2043}, \bibinfo{year}{2023}.
%Type = Misc
\bibitem[{Nations(2015)}]{SDG4}
\bibinfo{author}{U.~Nations}, \bibinfo{title}{{Sustainable Development Goal 4}, quality education}, \bibinfo{year}{2015}. \URLprefix \url{https://sdgs.un.org/goals/goal4}.
%Type = Article
\bibitem[{Poveda{-}Villal{\'{o}}n et~al.(2022)Poveda{-}Villal{\'{o}}n, Fern{\'{a}}ndez{-}Izquierdo, Fern{\'{a}}ndez{-}L{\'{o}}pez, and Garc{\'{\i}}a{-}Castro}]{DBLP:journals/eaai/Poveda-Villalon22}
\bibinfo{author}{M.~Poveda{-}Villal{\'{o}}n}, \bibinfo{author}{A.~Fern{\'{a}}ndez{-}Izquierdo}, \bibinfo{author}{M.~Fern{\'{a}}ndez{-}L{\'{o}}pez}, \bibinfo{author}{R.~Garc{\'{\i}}a{-}Castro},
\newblock \bibinfo{title}{{LOT:} an industrial oriented ontology engineering framework},
\newblock \bibinfo{journal}{Eng. Appl. Artif. Intell.} \bibinfo{volume}{111} (\bibinfo{year}{2022}) \bibinfo{pages}{104755}. \URLprefix \url{https://doi.org/10.1016/j.engappai.2022.104755}. \DOIprefix\doi{10.1016/J.ENGAPPAI.2022.104755}.
%Type = Article
\bibitem[{Ilkou et~al.(2023)Ilkou, Tolmachova, Fisichella, and Taibi}]{DBLP:journals/tlt/IlkouTFT23}
\bibinfo{author}{E.~Ilkou}, \bibinfo{author}{T.~Tolmachova}, \bibinfo{author}{M.~Fisichella}, \bibinfo{author}{D.~Taibi},
\newblock \bibinfo{title}{Collabgraph: {A} graph-based collaborative search summary visualization},
\newblock \bibinfo{journal}{{IEEE} Trans. Learn. Technol.} \bibinfo{volume}{16} (\bibinfo{year}{2023}) \bibinfo{pages}{382--398}. \URLprefix \url{https://doi.org/10.1109/TLT.2023.3242174}. \DOIprefix\doi{10.1109/TLT.2023.3242174}.
%Type = Inproceedings
\bibitem[{Ilkou and Koutraki(2020)}]{ilkou2020symbolic}
\bibinfo{author}{E.~Ilkou}, \bibinfo{author}{M.~Koutraki},
\newblock \bibinfo{title}{Symbolic vs sub-symbolic {AI} methods: Friends or enemies?},
\newblock in: \bibinfo{booktitle}{Proceedings of the {CIKM} 2020 Workshops co-located with 29th {ACM} International Conference on Information and Knowledge Management {(CIKM} 2020), Galway, Ireland, October 19-23, 2020}, volume \bibinfo{volume}{2699} of \textit{\bibinfo{series}{{CEUR} Workshop Proceedings}}, \bibinfo{publisher}{CEUR-WS.org}, \bibinfo{year}{2020}.
%Type = Inproceedings
\bibitem[{Li et~al.(2023)Li, Cheng, Zhang, Zhu, and Zhao}]{DBLP:conf/icalt/LiCZZZ23}
\bibinfo{author}{Z.~Li}, \bibinfo{author}{L.~Cheng}, \bibinfo{author}{C.~Zhang}, \bibinfo{author}{X.~Zhu}, \bibinfo{author}{H.~Zhao},
\newblock \bibinfo{title}{Multi-source education knowledge graph construction and fusion for college curricula},
\newblock in: \bibinfo{booktitle}{{IEEE} International Conference on Advanced Learning Technologies, {ICALT} 2023, Orem, UT, USA, July 10-13, 2023}, \bibinfo{publisher}{{IEEE}}, \bibinfo{year}{2023}, pp. \bibinfo{pages}{359--363}. \URLprefix \url{https://doi.org/10.1109/ICALT58122.2023.00111}. \DOIprefix\doi{10.1109/ICALT58122.2023.00111}.

\end{thebibliography}

% \appendix


% The sources for the ceur-art style are available via
% \begin{itemize}
% \item \href{https://github.com/yamadharma/ceurart}{GitHub},
% % \item \href{https://www.overleaf.com/project/5e76702c4acae70001d3bc87}{Overleaf},
% \item
%   \href{https://www.overleaf.com/latex/templates/template-for-submissions-to-ceur-workshop-proceedings-ceur-ws-dot-org/pkfscdkgkhcq}{Overleaf
%     template}.
% \end{itemize}

\end{document}